\documentstyle[preprint,aps]{revtex}
\tighten
\begin{document}

\newcommand{\bq}{\begin{equation}}
\newcommand{\eq}{\end{equation}}
\newcommand{\bqn}{\begin{eqnarray}}
\newcommand{\eqn}{\end{eqnarray}}
\newcommand{\nb}{\nonumber}
\newcommand{\lb}{\label}


\title {Spherical Self-Similar Solutions in Einstein-Multi-Scalar Gravity}
\author{Eric W. Hirschmann\thanks{Electronic Address: ehirsch@dirac.ph.utexas.edu } 
\ and Anzhong Wang\thanks{On leave from
 Departamento de F\' {\i}sica Te\' orica,
Universidade do Estado do Rio de Janeiro, Cep.
20550-013, Rio de Janeiro~--~RJ, Brazil. Electronic Address: wang@symbcomp.uerj.br}\\
\small Center for Relativity, The University of Texas, Austin, Texas, 78712-1081}

\maketitle

\begin{abstract}
We consider a general non-linear sigma model coupled to Einstein gravity and
show that in spherical symmetry and for a simple realization of self-similarity,
the spacetime can be completely determined.  We also examine some more specific
matter models and discuss their relation to critical collapse.   

\end{abstract}

\newpage

\section*{Introduction}
The problem of gravitational collapse has received a great deal of attention
in the past several years.  This is due in large part to Choptuik's results
on the collapse of a real scalar field coupled to gravity \cite{MWC1993}.  
Choptuik showed that
the strong field regime exhibited rather striking and unexpected behavior.
This included the surprising result that black holes can form 
with arbitrarily small mass and that a scaling relation exists for the 
black hole mass.  In addition, the ``critical" solution exactly at 
the threshold of black
hole formation possesses a discrete self-similarity and serves as an intermediate
attractor for near-critical solutions in the space of initial data.   

Considerable effort has been made to understand these results from both 
computational and analytic perspectives.  The general behavior is now rather
well understood and a very nice review article summarizes the relevant 
literature and current state of the art \cite{G1997}.  
However, as pointed out there, progress in understanding critical phenomena
from a purely analytic perspective
has been slower in coming.  This is largely due to the well known  
difficulties associated with finding closed form solutions to Einstein's 
equations in dynamical situations.  One of the few closed form solutions to the Einstein
equations which appears somewhat related to critical phenomena is the so-called
Roberts solution.  It had been written down several times in the 
literature \cite{MC}\cite{R1989},
but was rediscovered in an attempt to get an analytic grasp of the features
seen in critical collapse \cite{B1994}.  
The solution is spherically symmetric and   
continuously self-similar.  It also contains timelike singularities.  
When these singularities are cut out and replaced with Minkowski space,
the spacetime describes a collapsing ball of scalar field.  In addition, as a parameter
is varied, the solution describes dispersal of the field to infinity or the formation
of an apparent horizon. 

Our attempt here is to extend this solution to a wider class of 
matter models.  Numerical work has shown that critical behavior arises in
a variety of matter models coupled to gravity.  
Thus one might expect the Roberts solution 
to be a member of a larger class of these ``toy models."  

Indeed, we find that the basic spacetime for scalar field collapse as 
described by the Roberts solution turns out to extend virtually 
unchanged to a large class of massless interacting scalar fields.  It
is this result that we present in the next section and follow with
some specific examples.

\section*{The Model and the Spacetime Solution}

We are interested in considering a generalization of the Einstein-scalar
field problem.  So we turn our attention to Einstein gravity coupled to 
multiple interacting scalar fields.  The particular form of the theory
we will consider is given by the action
\bq
\lb{eq1}
S = \int d^{4}x \sqrt{ - g} \left\{ R + 2 L_{m}\right\},
\eq
where the Lagrangian density for the matter fields is given by
\bq
\lb{eq2}
L_m = - \frac{\lambda}{2} \, G_{AB}(\phi^C) \phi^{A}{}_{,\alpha} \phi^{B}{}^{,\alpha} 
\eq
where we have $N$ scalar fields $\phi^{A}$ ($A=1\ldots N$) and $G_{AB}$ is a set of functions of the
scalar fields which must be specified to fix the model.  In general, the functions making up $G_{AB}$ 
take the form of a Riemannian
metric on the $N$-dimensional internal space of the scalar fields (the target space).  
Thus the $N$ real scalar fields, $\phi^{A}$, are coupled to Einstein gravity with $\lambda$ 
a dimensionless coupling constant.  In fact, any parameters entering the target space metric $G_{AB}$
will also be dimensionless and this in turn will allow solutions to be self-similar. 
The theory defined by Eq.(\ref{eq2}) is thus a particular 
case of more general Einstein-multi-scalar theories of gravity such as those discussed in \cite{DF1992}.

Varying the action with respect to the metric and the matter fields  
yields the Einstein field equations and the equations of motion for the matter:
\bqn 
\lb{eq3}
R_{\mu\nu} & = & \lambda \, G_{AB}(\phi^C) \, \phi^{A}{}_{,\mu} \, \phi^{B}{}_{,\nu}, \\ 
\lb{eq4}
\phi^{A}{}_{,\mu}{}^{;\mu} & = & - \phi^{B}{}_{,\mu} \, \phi^{C}{}^{,\mu} \, \Gamma^{A}_{BC},   
\eqn 
where, as usual, a comma and a semicolon denote partial and covariant 
differentiation, respectively.  We also define the Christoffel symbols on the
target space
\bq
\lb{eq5}
\Gamma^{A}_{BC} =  {1\over2} \, G^{AD} \left[ 
	{ \partial G_{BD} \over \partial \phi^{C} } 
      + { \partial G_{CD} \over \partial \phi^{B} } 
      - { \partial G_{BC} \over \partial \phi^{D} } 
						\right] .
\eq
Because of the contracted Bianchi identities $T_{\mu\nu;\lambda} \, g^{\nu\lambda} = 0$, 
Eqs.(\ref{eq3}) - (\ref{eq4}) are not independent. 
In the following, we shall
use this fact to solve these equations by properly choosing the independent ones.

Working in spherical symmetry, we can write the metric in double null coordinates 
\bq
\lb{eq6}
ds^{2} = - 2 e^{2\sigma}dudv + r^{2}\left(d\theta^{2} + \sin^{2}\theta d\varphi^{2}\right),
\eq
where $\sigma$ and $r$ are functions of the two null coordinates $u$ and $v$ only. The metric
is invariant under the coordinate transformations $ \bar{u} = \bar{u}(u),\; \bar{v} = 
\bar{v}(v)$. In addition to assuming spherical symmetry, we are interested in spacetimes 
that are continuously self-similar (CSS).  Stated invariantly, we are assuming the existence of
a homothetic Killing vector field $\xi$ satisfying \cite{CT1971}    
\bq
\lb{eq7}
{\cal L}_{\xi} g_{\mu\nu} = \xi_{\mu;\nu} + \xi_{\nu;\mu} = 2 g_{\mu\nu},
\eq
where ${\cal L}$ denotes the Lie derivative.  It is then straightforward to show
that the homothetic vector field is
\bq
\lb{eq8}
\xi = u \frac{\partial}{\partial u} + v \frac{\partial}{\partial v}.
\eq
In addition, it follows that the the metric coefficients take the form
\bq
\lb{eq9}
\sigma  =  \sigma(z), \qquad r = u s(z),
\eq 
where $z \equiv v/u$.    
We also need to know the form of the scalar fields in the presence 
of the continuous self-similarity.
The simplest ansatz is to assume 
\bq
\lb{eq10}
{\cal L}_\xi \phi^A = 0 
\eq 
such that the scalar fields will depend only on the coordinate $z = v/u$ as with the
metric coefficients.  This is, in fact, the ansatz we will make here, but it is not
necessarily the most general.  Depending on the symmetries of the $N$-dimensional target
space defined by the scalar fields, one possible generalization is to allow the homothetic motion to be 
accompanied by a change in the scalar fields which respects the target space symmetries.  
For example, in the case of a complex scalar field ($N = 2$ and $G_{AB} = \eta_{AB}$, 
a flat target space), there is a $U(1)$ 
symmetry which could be incorporated into the self-similarity ansatz for the fields \cite{HE1997}.
Another possible generalization is to make a kind of hedgehog ansatz for the scalar fields
before imposing self-similarity.  In
this case, the $\phi^{A}$ fields will have some nontrivial angular dependance.  An example is 
the self-similar solution of a spherically symmetric $O(4)$ non-linear sigma model which 
models the relativistic collapse of a global texture \cite{RD1992}. 

 For the ansatz given by Eg.(\ref{eq10}), 
the Einstein field equations (\ref{eq3}) take the form
\bqn
\lb{eq11}
2z\sigma'(s - z s') + z^{2}s'' &=&  - \frac{\lambda z^{2}s}{2} G_{AB} \phi^{A}{}' \phi^{B}{}',\\
\lb{eq12}
2\sigma' s' - s'' &=&  \frac{\lambda s}{2} G_{AB} \phi^{A}{}' \phi^{B}{}',\\
\lb{eq13}
2(\sigma' + z \sigma'') + 2z \frac{s''}{s} &=&  - \lambda z G_{AB} \phi^{A}{}' \phi^{B}{}',\\
\lb{eq14}
\left[\frac{(s^{2})'}{z}\right]^{'} &=&  \frac{e^{2\sigma}}{z^{2}}.
\eqn
where a prime indicates differentiation with respect to the similarity variable $z$.
 From Eqs.(\ref{eq11}) and (\ref{eq12}) we find that $\sigma = const$. By properly 
rescaling the null coordinates $u$ and $v$, we can set $\sigma = 0$. Thus, without loss of 
generality, in the following we shall consider only the case where $\sigma = 0$. Substituting into 
Eq.(\ref{eq14}) we find that
\bq
\lb{eq15}
s^{2}(z) = \alpha z^{2} - z + \beta,
\eq
where $\alpha$ and $\beta$ are two integration constants. Meanwhile, Eqs.(\ref{eq11})
- (\ref{eq13}) reduce to the single equation
\bq
\lb{eq16}
G_{AB} \phi^{A}{}' \phi^{B}{}' = \frac{\gamma}{s^{4}},
\eq
where $\gamma \equiv (1 - 4\alpha\beta)/(2\lambda)$. It is interesting to note that without having specified the 
target space metric $G_{AB}$, we have been able to completely determine the metric coefficients.  
Thus, spherical symmetry and our simple ansatz for self-similarity completely fix the 
spacetime geometry to be the same for all of these non-linear sigma models. 
In particular, the spacetime and its global structure 
is the same as that of the Roberts solution. \footnote{Strictly speaking, 
this will be true provided $1-4\alpha\beta>0$ as in the Roberts solutions. 
However, in the following, we will encounter a situation where this is
not necessarily the case.}

\section*{Solutions of the Matter Equations}

To complete the solution, we need to specify the non-linear sigma model 
by choosing a form for $G_{AB}$.  In the following we will restrict
ourselves to a 2-dimensional target space of the form 
\bqn 
\lb{eq16a} 
G_{AB} & = & \pmatrix{ 1 & 0 \cr 0 & f(\phi) } 
\eqn 
where, for clarity, we define $\phi^1 = \phi$ and $\phi^2 = \psi$.

The matter field equations Eq.(\ref{eq4}) can now be written as 
\bqn  
\lb{eq17}
\left[s^2 \phi'\right]' & = & {s^2\over2}  f_{,\phi}(\phi) \, \psi'{}^2 \\
\lb{eq18}
\left[s^{2}f(\phi) \psi'\right]^{'} & = & 0.
\eqn  
Clearly, the second equation has the first integal,
\bq
\lb{eq19}
\frac{d\psi}{dz} = \frac{c_{0}}{f(\phi) s^{2}},
\eq
where $c_{0}$ is an integration constant. Inserting the above equation into Eq.(\ref{eq16}),
we obtain
\bq
\lb{eq20}
\frac{d\phi}{dz} = \pm \left[\gamma - \frac{c_{0}^{2}}{f(\phi)}\right]^{1/2}\frac{1}{s^{2}}.
\eq
To solve Eqs.(\ref{eq19}) and (\ref{eq20}),
we need to specify the function $f(\phi)$. In the following, we shall consider some
particular cases.

\bigskip

\noindent {\bf I} \quad $f(\phi) = e^{-\phi}$ 

This case corresponds to the non-linear $\sigma$-model considered
in \cite{HE1997} for the case $\kappa > 0$. In fact, if we set
\bq
\lb{eq21}
F = \frac{1}{\sqrt{\kappa}}\frac{1 + i\tau}{1 - i\tau},\qquad \tau = \frac{1}{4}\left(\psi +
2ie^{\phi/2}\right),
\eq
then  the Lagrangian density of the complex matter field $F$ takes the form
\bq
\lb{eq22}
 L_{m} = - \frac{\left|\nabla F\right|^{2}}{(1 - \kappa |F|^{2})^{2}} = -\frac{1}{16\kappa}
\left(\phi^{,\alpha}\phi_{,\alpha} + e^{-\phi}\psi^{,\alpha}\psi_{,\alpha}\right),
\eq
which corresponds to the one given by Eqs.(\ref{eq2}) and (\ref{eq16a}) with
\bq
\lb{eq23}
\lambda = \frac{1}{8\kappa},\qquad f(\phi) = e^{-\phi}.
\eq
As noticed in \cite{HE1997}, this also corresponds to the Brans-Dicke (BD) theory of gravity considered in
\cite{LC1996} with 
\bq
\lb{eq24}
\lambda = \omega + \frac{3}{2},
\eq
where $\omega$ is the BD coupling constant. Note the difference between $\lambda$ used here and the one used
in \cite{LC1996}.

Substituting Eq.(\ref{eq23}) into Eqs.(\ref{eq19}) and (\ref{eq20}), and then integrating them, we obtain    
\bqn
\lb{eq25}
\phi & = & 2 \ln\frac{2g}{1 + g^{2}} + \ln {\gamma\over c_{0}{}^{2}} ,\nb\\
\psi & = & \mp {\sqrt{\gamma} \over c_0 } \frac{4}{1 + g^{2}} + \psi_{0},
\eqn
where
\bqn
\lb{eq26}
g & = & c_{1}\left(\frac{z - z_{+}}{z - z_{-}}\right)^{\pm \sqrt{\kappa}},\nb\\
z_{\pm} & = & \frac{ 1 \pm \sqrt{1 - 4 \alpha\beta}}{2\alpha},
\eqn
and  $c_{1},\; \phi_{0}$ and $\psi_{0}$ are integration constants. Note that in this case the parameters $\alpha$
and $\beta$ have to be chosen such that $\alpha\beta < 1/4$. Otherwise, no solutions exist.

\bigskip

\noindent {\bf II} \quad $f(\phi) = \sin^{2}2\phi$

For this case we have the non-linear $\sigma$-model considered in \cite{HE1997}
for $\kappa < 0$. Actually, if we set
\bq
\lb{eq27}
F = \frac{\tan \phi}{\sqrt{|\kappa|}}\left(\sin2\psi + i\cos2\psi\right),
\eq  
we find that
\bq
\lb{eq28}
L_{m} = - \frac{|\nabla F|^{2}}{(1 - \kappa|F|^{2})^{2}} = -  \frac{1}{|\kappa|}\left(\phi^{,\alpha}
\phi_{,\alpha} + \sin^{2}2\phi\psi^{,\alpha}\psi_{,\alpha}\right),
\eq
which corresponds to Eqs.(\ref{eq2}) and (\ref{eq16a}) with
\bq
\lb{eq29}
\lambda = \frac{2}{|\kappa|},\qquad f(\phi) = \sin^{2}2\phi.
\eq  
Inserting the above expressions into Eqs.(\ref{eq19}) and (\ref{eq20}), we find the following solution 
\bqn
\lb{eq30}
\cos2\phi & =& \left[1 - \frac{2\lambda c_{0}^{2}}{1 - 4\alpha\beta}\right]^{1/2}\cos g,\nb\\
\psi &=& \pm \frac{1}{2}\tan^{-1}\left[\frac{\sqrt{\gamma}}{c_{0}} \tan g\right] + \psi_{0},
\eqn
but now 
\bq
\lb{eq31}
g \equiv \pm \sqrt{|\kappa|}\ln\left(\frac{z - z_{+}}{z - z_{-}}\right) + c_{1},
\eq
where $z_{\pm}$ are given by Eq.(\ref{eq26}), and $c_{1}$ and $\psi_{0}$ are other
 integration constants. Similar to
the last case, now the condition $\alpha\beta < 1/4$ has to hold.

Note that when $\kappa = 0$, the self-similar solution of the non-linear $\sigma$-model of Eq.(\ref{eq22}) or Eq.(\ref{eq28})
with the ansatz (\ref{eq10}) is the same as that of Roberts \cite{R1989}.
As a matter of fact, in the latter case we have $f(\phi) = 1$. Then, from Eqs. (\ref{eq19}) and (\ref{eq20})
we find that
$\psi = A\phi$, where $A$ is a constant.

\bigskip

\noindent {\bf III} \quad $f(\phi) = \frac{1}{\lambda}e^{-\phi},\; 
\lambda = {1\over2}\left(3 + 2\omega\right)$

Finally, we consider the case corresponding to the BD theory
considered in \cite{LC1996}. As noted previously, when $\omega > - 3/2$, 
it is equivalent to the non-linear $\sigma$-models
with $\kappa > 0$. However, when $\omega < - 3/2$, such a
correspondence no longer exists. Therefore, in the following we shall
consider only the case where $\omega < - 3/2$. Before proceeding and by way 
of justification, we note that the case where $\omega < 0$ is also worthy
of study for more than its intrinsic theoretical interest.  In particular, 
for sufficiently negative $\omega$, the theory passes all 
the experimental and observational constraints as  
does the case of positive $\omega$ \cite{W1993}. However, when
$\omega < -3/2$, the static spherical spacetimes are quite different
from the ones with $\omega > - 3/2$. In particular, in
the former case ``cold" spherical black holes exist \cite{CL1993}, while 
in the latter case, the only static spherical black hole
is the Schwarzschild black hole \cite{H1972}.

Inserting the above expressions for $\lambda$ and $f(\phi)$
 into Eqs.(\ref{eq19}) and (\ref{eq20}), we find that
solutions exist for both $\alpha\beta < 1/4$ and $\alpha\beta > 1/4$.  

\medskip
\noindent $\quad$ {\bf III}a \quad $1 - 4\alpha\beta > 0$

In this case, we have
\bq
\lb{eq32}
\phi =  \ln\left[\cos^{2}g\right] + \phi_{0},\quad \psi = c_{0}|3 + 2\omega|\tan g + \psi_{0},
\eq
where
\bq
\lb{eq33}
g \equiv  \frac{\epsilon}{(8\lambda^{2}c_{0}^{2})^{1/2}}\ln\left(\frac{z - z_{+}}{z - z_{-}}\right) + c_{1},
\eq
and $z_{\pm}$ are given by Eq.(\ref{eq26}), and $\epsilon = sign(\alpha)$.

\medskip
\noindent \quad {\bf III}b \quad $1 - 4\alpha\beta < 0$

Now we have
\bqn
\lb{eq34}
\phi & = & - \ln\left[\sinh^{2}\left(- \frac{g}{2}\right)\right] + \phi_{0},\nb\\
\psi & =& \pm \frac{2(4\alpha\beta - 1)}{c_{0}|2\omega + 3|}\mbox{cotanh}\left(- \frac{g}{2}\right) + \psi_{0},
\eqn
where
\bq
\lb{eq35}
g \equiv \pm \left|\frac{2c_{0}^{2}(2\omega + 3)}{4\alpha\beta -1}\right|^{1/2}\tan^{-1}\left[
\frac{2\alpha z - 1}{(4\alpha\beta -1 )^{1/2}}\right] + c_{1}.
\eq
For  $r$ to be real, we need to impose the condition $\alpha > 0$ and $\beta > 0$ in this case. 

Clearly, by giving different $f(\phi)$, we can obtain various solutions for the matter fields $\phi$ and $\psi$. 
However, these solutions are not all independent. For some particular choices, they may correspond to the same  
matter fields. This follows, of course, from the diffeomorphism invariance
of the target space manifold.  For example, the choice 
$f(\phi) = \sinh^{2}2\phi$ gives the same solution as that given in
Case {\bf I}. This is closely related to the fact that the matter fields described by the complex function $F$
has a $SL(2, R)$ symmetry for the case $\kappa > 0$ \cite{HE1997}. 
But for $\kappa < 0$, no such symmetry exists.
We would thus expect that the solution given in Case {\bf II} is unique.

Since the spacetime geometry is independent of the particular matter model, 
the global structure of the
spacetimes should be the same as that of the Roberts solution \cite{B1994}, provided that $1 - 4\alpha\beta > 0$.  
Of course, the Ricci scalar and the mass will also be unchanged  
\bq
\lb{eq36}
R = { uv \over r^4 } ( 1 - 4 \alpha\beta ) \qquad M = - {uv \over 4r} ( 1 - 4 \alpha\beta ) .
\eq
Again, one can remove the timelike singularities which are present
and replace them with Minkowski space, matching at the advanced time $v=0$.  
In this way, we imagine the scalar fields to be turned on at $v=0$.
Thus, the solution with $\beta = 1$ resemble critical collapse in the sense that
for $\alpha < 0$ black holes 
are formed (supercritical regime), while for the case $0 < \alpha < 1/4$, 
no black holes are formed and the fields disperse to infinity (subcritical regime) \cite{B1994}.  
Note, however, that because of the matching, the solutions are $C^0$ in contrast
to the smooth, attracting critical solutions which have been found in
numerical calculations. 

When $1 - 4 \alpha\beta < 0$, which is the case {\bf IIIb}, there is no correspondence
to the Roberts solution. From Eq.(\ref{eq36}) we can see that to have the mass be non-negative, the solution
has to be restricted to the regions $u > 0,\; v > 0$ and $u < 0,\; v < 0$, which will be referred, respectively, as
regions $I$ and $II$. In region $II$ the solution can be considered as representing the gravitational 
collapse of two scalar fields and  a point-like singularity is finally formed at $u = 0 = v$. 
This singularity is direction-dependent. When the singularity is approached along different directions, 
the singular behavior will be different. In particular, if one approaches the singularity 
along the  null hypersurface $u = 0$ or the one 
$v = 0$, everything seems regular. On the
hypersurfaces $u = 0, \; v < 0$ and $v = 0,\; u < 0$, the two scalar fields carry null data. Using
the arguments for  
the case $1 - 4 \alpha\beta > 0$ 
\cite{B1994}, we can replace  the solution in  the regions $u > 0,\; v < 0$ and $u < 0,\; v > 0$
by two flat regions. Once this is done, the spacetime needs to be further extended
beyond the hypersurfaces $v = 0, \; u > 0$ and $u = 0,\; v > 0$, which are 
 Cauchy horizons. The extension beyond them are not unique. For example, one can match the two flat regions
to the solution given by Eq.(\ref{eq34})  for $u > 0,\; v > 0$, or make an analytic 
extension and replace region $I$ by another flat region. However,
in either case the singularity at $u = 0 = v$ is naked.  Figure 1 gives a 
conformal diagram of the spacetime summarizing the above results.   

Finally, we would like to note that the conclusion that the spacetime 
geometry is independent of the choice of the particular scalar field model 
is based on the fact that the spacetime realizes a fairly simple assumption
of self-similarity. Otherwise, it is easy to show that this
conclusion does not hold.  Therefore, for the general perturbations 
(which do not have such a symmetry), it would be expected that
they will depend on the choice of non-linear sigma model. In particular, 
there are no {\em a priori } reasons to expect that the spectrum of
the unstable modes of the perturbations of the solution $\alpha = 0$ 
and $\beta = 1$ for a general choice of  $G_{AB}$ 
is the same as that for the particular Roberts solution 
as was considered in \cite{F1997}.  
This point deserves further investigation and will 
be considered in future work.

\section*{Acknowledgment}

We would like to thank M.W. Choptuik and S.L. Liebling for valuable discussions. Part of this work was done
while one of the authors (AW) was a visitor at the Center for Relativity, the University of Texas at Austin. He
would like to express his gratitude to the Department for hospitality. The financial assistance from CNPq, 
FAPERJ, and UERJ  (AW) is gratefully 
acknowledged.  EWH was supported in part by the National Science Foundation
under Grant Nos.\ PHY93-10083 and PHY93-18152\@.



\pagebreak


\begin{figure}
\caption{The conformal diagram for the spacetime described by 
case {\bf III}b.  The singularity is at $(u,v) = (0,0)$ and the 
Cauchy horizons are at $u=0, v>0$ and $v=0, u>0$.  Region $I$ can
be replaced either (analytically) by flat space or (continuously)
by the solution given by Eq.(\ref{eq34}).  For more details, see   
the text.
}
\label{fig1}
\end{figure}

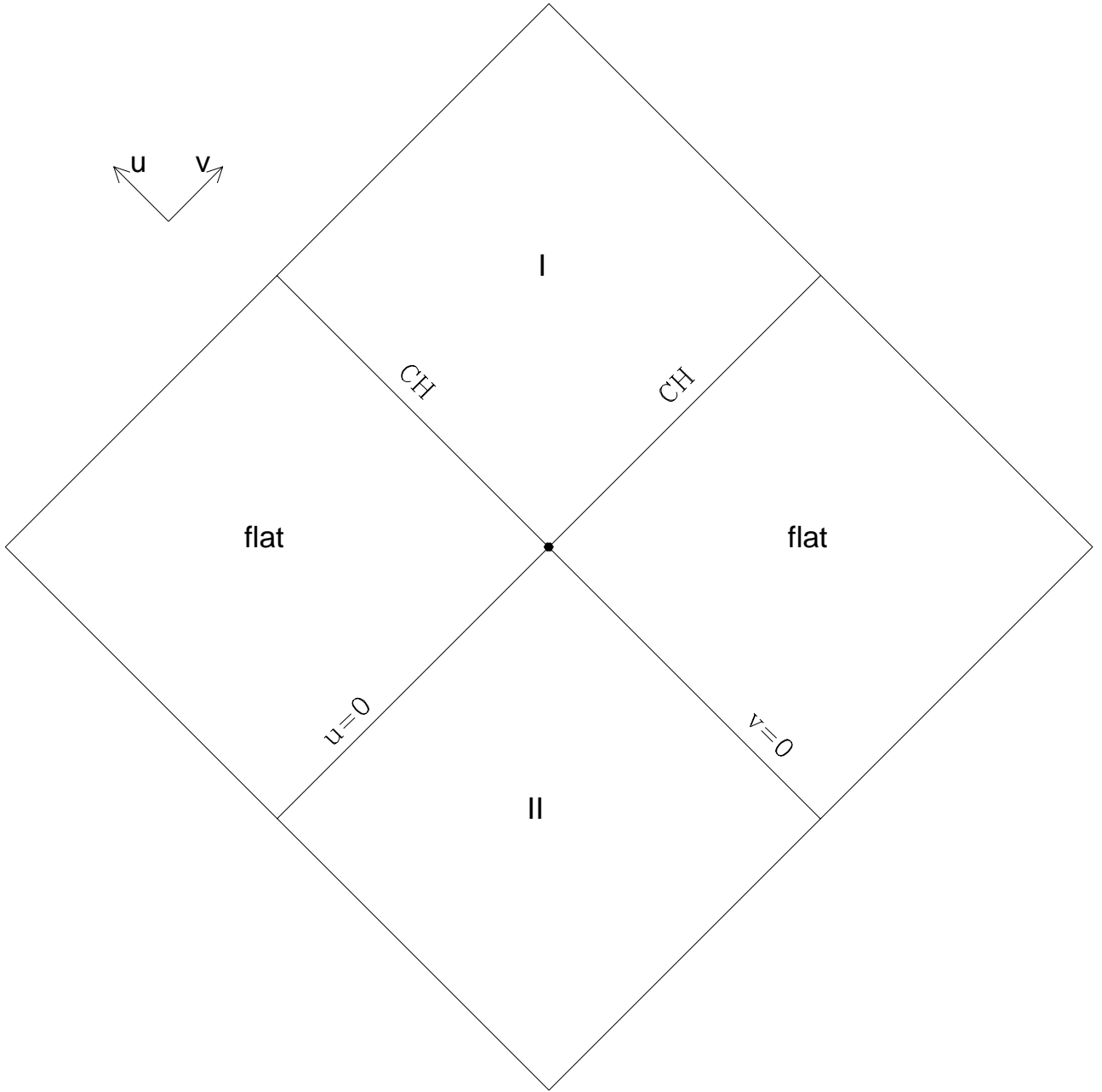
\end{document}